\documentclass{collective-intelligence}

\usepackage{graphicx}
\usepackage{comment}

\begin{document}
\bibliographystyle{agsm}

\title{CROWD MEMORY: LEARNING IN THE COLLECTIVE}
%
%
%
%
%

\numberofauthors{4}
%
\author{
%
%
\\
\parbox[t]{16cm}{\centering
Walter S. Lasecki, Samuel C. White, Kyle I. Murray, and Jeffrey P. Bigham\\
      \affaddr{University of Rochester Department of Computer Science}\\
      \affaddr{ ROC HCI Lab}\\
      \affaddr{160 Trustee Rd., Rochester, NY. 14627}\\
      \email{\{wlasecki,jbigham\}@cs.rochester.edu\\
      \{samuel.white,kyle.murray\}@rochester.edu}
}
}

\maketitle
\begin{abstract}
  Crowd algorithms often assume workers are inexperienced
  and thus fail to adapt as workers in the crowd learn a task. These
  assumptions fundamentally limit the types of tasks that 
  systems based on such algorithms can handle. This paper
  explores how the crowd learns and remembers over time
  in the context of human computation, and how more realistic
  assumptions of worker experience may be used when designing
  new systems.  We first demonstrate that the crowd can recall
  information over time and discuss possible implications of
  crowd memory in the design of crowd algorithms. We then
  explore crowd learning during a continuous
  control task. Recent systems are able to disguise dynamic groups of
  workers as \emph{crowd agents} to support continuous
  tasks, but have not yet considered how such agents are able to
  learn over time. We show, using a real-time gaming setting, that
  crowd agents can learn over time, and `remember' by passing
  strategies from one generation of workers to the next, despite
  high turnover rates in the workers comprising them. We
  conclude with a discussion of future research directions for
  crowd memory and learning.


\end{abstract}

\section{INTRODUCTION}


Crowdsourcing has shown that \emph{the crowd}, a dynamic pool of workers
of varying reliability, can be effectively used
for a variety of computational tasks. In this paper, we explore how crowds
collectively learn and recall information over time in the context of human
computation. A better understanding of crowd memory may help enable the crowd
to complete tasks that require domain knowledge, experience, or a consistent
strategy.

The dynamic nature of the crowd means that workers come and go, and no
specific worker can be relied upon to be available at a given time or
to continue working on a job for a set amount of time.
Existing systems take advantage of a worker's prior learning to provide
knowledge and understanding of the world and improve their ability
to complete a task, but most do not consider
learning the task itself.  Understanding crowd memory may also enable us
to better design systems that manage memory expectations - for instance,
formulating guarantees about independence in tasks that require it (such
as voting or user trials). In general, we assume that interactions between
crowd workers happen only when mediated by the crowdsourcing system,
e.g. crowd workers do not learn from one another after meeting in a
coffee shop. As a
result, the collective memory at a point in time is a function of the
union of the individual memories of all active workers connected at that time.

We begin by demonstrating that crowds can remember information
over periods of time. This basic form of crowd memory relies on workers
remembering attributes of a task beyond their submission, and re-recruiting
the same workers for subsequent tasks. This concept illustrated with an
experiment on Mechanical Turk in which knowledge is introduced via a task, 
then the crowd is queried at intervals ranging from 0 to 12 hours.
Over this time span, our ability to attract the same workers
diminishes and each worker's memory fades, therefore so does the observed
crowd memory. 


We then investigate collective learning in the context of a continuous control
task (navigation) in Legion \cite{legion}. Legion presented a
paradigm in which a crowd of workers acts as a
single more reliable agent. However, it is not clear that such crowds can
simulate the memory that a single worker would usually have for previous
events occuring in a task. This type of memory in a collective with high
turnover is seen in organizational learning in which knowledge is
passed between members, allowing the group to maintain knowledge
beyond the tenure of any individual.
Using this model, we test the ability
of the crowd to implicitly pass knowledge to newer members through
\emph{demonstration points}, events that provide feedback to the crowd and
demonstrate specific prior knowledge. We then discuss how these
demonstration points can be used, and potential ways to enable transfer
learning through explicit communication.


\section{RELATED WORK}
We ground our exploration of crowd memory and learning in two areas:
(i) the crowd-powered systems that motivate our work on crowd memory
and learning, and (ii) sociological theories of organizational learning.

\subsection{Crowd-Powered Systems}

A number of crowd-powered systems have been developed that engage workers
in multiple tasks with the potential for learning. For instance, games
with a purpose often reward workers with points that accumulate over
time \cite{gwap}, e.g. the ESP Game \cite{espgame}, FoldIt \cite{foldit}.
Systems like Shepherd aid learning by giving workers explicit feedback on
their work \cite{shepherd}, whereas other systems build in an implicit
learning phase. For instance, VizWiz asks crowd workers to complete old
tasks as they wait for new work to arrive \cite{vizwiz}. Many crowd
systems are iterative, potentially allowing crowd workers to contribute
over multiple similar or related tasks \cite{turkit,crowdforge}. Systems
like Legion \cite{legion} engage workers in continuous tasks potentially
allowing crowd workers to learn from one another interactively. Finally,
most crowd-powered systems issue multiple similar
jobs as they are applied to new instances of problems, providing an
opportunity for crowd workers to learn how to do those jobs.

Many sources of crowd labor build in mechanisms designed to encourage
learning. Mechanical Turk offers qualifications that crowd workers can
complete in order to be eligible for certain tasks. At a lower level,
Mechanical Turk also provides a unique identifier for each crowd worker
that enables systems to identify (and re-identify) workers. Such identifiers
allow systems to estimate whether workers are already familiar with a task.
However, in general, it can be difficult to efficiently recruit workers that
a system has seen before in current crowd marketplaces.

\subsection{Organizational Learning}

Our understanding of crowd memory and learning in continuous real-time
 and collective answer tasks is based on organizational
learning theory (summarized in \cite{organizational-learning}). From this
perspective, crowd learning happens as the collective experience is
synthesized into routines (strategies and conventions) that guide future
behavior. Routines encode the appropriateness or legitimacy of different
actions, and are viewed as forming over time from experience. Crowd action
is directed toward a target, and it is the synthesis of this target and
collection of routines that results in a final decision on how to act.

For instance, in the case of collectively navigating a maze within Legion,
a crowd is provided with their target via the task instructions, e.g.
``complete the maze.''  They may learn over time that turning left at
every intersection helps them make progress most quickly and encapsulate
this into a strategy (routine) that is continued on by the crowd even
after specific experiential knowledge of this rational is lost due to
crowd turnover.

Our working assumption is that many of the processes outlined in the
context of organizational learning also apply to crowdsourcing repeated
or continuous tasks. As with much of the work in crowdsourcing, the
general structure seems to mirror existing work, but may differ in
important ways. For instance, much of the research characterizing
organizational learning was conducted with traditional organizations,
e.g. business or university classes, which operate over different time
scales than existing crowd markets and may have formal processes
designed to support peer learning.

\section{CROWD MEMORY}
 
 Current crowdsourcing systems assume that workers do not retain
 task knowledge after its completion. This assumption is
 designed to best accommodate new workers, but also accounts
 for the fact that workers often get compensated very little for their
 effort, and make up for this by completing large numbers of tasks
 each day. With such high volume, workers seemingly cannot
 be expected to remember aspects of a single task. We show that this
 is not always the case, and that workers returning to a
 task are able to leverage prior experience.

 \begin{figure}
\includegraphics[width=20.1pc]{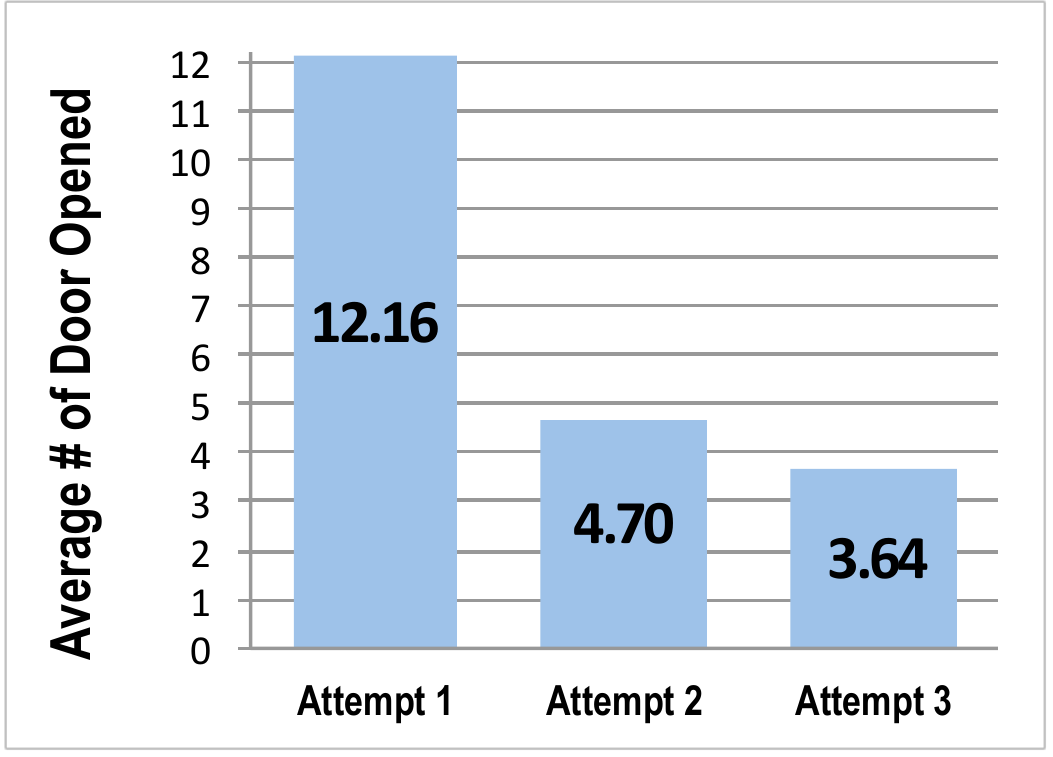}
\vspace{-1pc}
\caption{Average number of doors explored during a workers' first, second, and
		  third attempts at the task. The average quickly converges to near
		  the optimal value of three.}
\label{fig:nrt_fig}
\end{figure}

\subsection{Experiments}
 To test whether or not members of the crowd remember aspects of
 a task beyond its submission we presented workers with a
 task containing a specific pattern. We used a Javascript
 game that displays set of nine doors in a grid pattern, with gold
 behind one of them (clicking revealed what was behind a door).
 When the worker found the gold, the round was completed. To ensure
 workers were able to learn the pattern within a single
 session, workers had to complete three rounds (making 3 the minimum
 number of doors that had to be explored in a task). We fixed the position of
 the gold so that it always rests behind the same door.
 
 The same task was then posted again at a later time, allowing us to measure the difference in the
 number of doors which had to be explored in subsequent sessions. To test the
 role of time in memory, we vary the interval between postings of the task from 0 to 12 hours.

\subsection{Results}

 We found that, on average, workers were very quick to pick up on this simple
 pattern, with a majority of them learning it perfectly within the first 2
 rounds. Figure \ref{fig:nrt_fig} shows the average number of doors checked on
 the first, second, and third attempts to solve the problem. Workers had a
 median count of 12 doors opened on the first try, and 3 (optimal)
 for the second and third attempts.
 
 Workers were capable of remembering these tasks very well. Once most workers
 learned the pattern behind the task, they retained the information for
 the rest of the day. We tested over a 12 hour span and had individual
 times between tasks of up to 6 hours. Workers answering questions
 1 hour after their previous answer averaged an improvement
 of 2.78 doors, 76.8\% of the 3.34 doors for those answering immediately after.
 Workers who answered 6 hours after their previous completion of the task
 showed an average improvement of 2.75 fewer doors, 99\%
 that of worker answering questions only an hour later.

\subsection{Discussion}

 These results show that workers who do remember task information
 beyond its completion are not significantly effected by the time,
 on a within-day basis.
 The ability to store knowledge and skills in a crowd of workers is
 critical to many real-time and nearly real-time tasks. For example,
 VizWiz is an application which asks workers to answer audio
 questions based on visual information. Trained crowds are
 important for this task in two ways. First, questions are often
 submitted with only partial or no audio recorded. In this case,
 an experienced worker may be able to infer what the question
 will be based on previous experience and understanding
 the goal of the task, however, a worker with no experience may be
 uncertain of what to do in the task since they were not provided with
 the information they expected from instructions. Thus they may not
 be able infer the correct action to take, and requiring users to resubmit
 their question. Second, images submitted do not always accurately
 capture the information needed to answer the question, such as when a label
 is only partially viewable. Workers can submit directions in response in
 order to instruct users as to how the image angle can be adjusted to better
 capture the information. If the user submits two parts of the same label,
 it is possible for a single person, remembering prior submissions, to
 answer such questions. If we remove the ability to remember or
 assume independent workers, there will be a significantly higher number
 of expected re-submissions.
 
 Designing systems specifically to leverage
 worker memory may lead to improved performance. For example,
 directing subsequent VizWiz questions
 that occur over short time spans to similar groups of workers may help
 increase the retention rate of workers and decrease the number of
 submissions required to get a correct answer, reducing the average
 response latency.
  Demonstration tasks could also be provided to pre-recruited workers
  (such as those used by quikTurkit or Adrenaline \cite{adrenaline}),
  in order to train sets of workers while they wait for a task to begin.


\vspace{1pc}
\section{LEARNING FROM THE COLLECTIVE}

Not all tasks can be cleanly divided and solved separately, or in
an offline manner. Some problems must to be solved as a single
task, in real-time. Recent crowdsourcing systems have introduced
a model of the crowd in which the collective acts as a single worker
(or \emph{crowd agent}) that is capable of these continuous
real-time tasks \cite{legion}.
While this approach is effective in a variety of domains,
questions remain as to the properties of such agents. When
a single user is performing a task, such as operating an interface,
they can remember their previous actions and states of the system.
We show that a crowd agent
can also remember information is has learned previously by having
longer-serving members of the crowd demonstrate that knowledge
implicitly via their actions. 

\begin{figure}
\centering
\includegraphics[width=20.1pc]{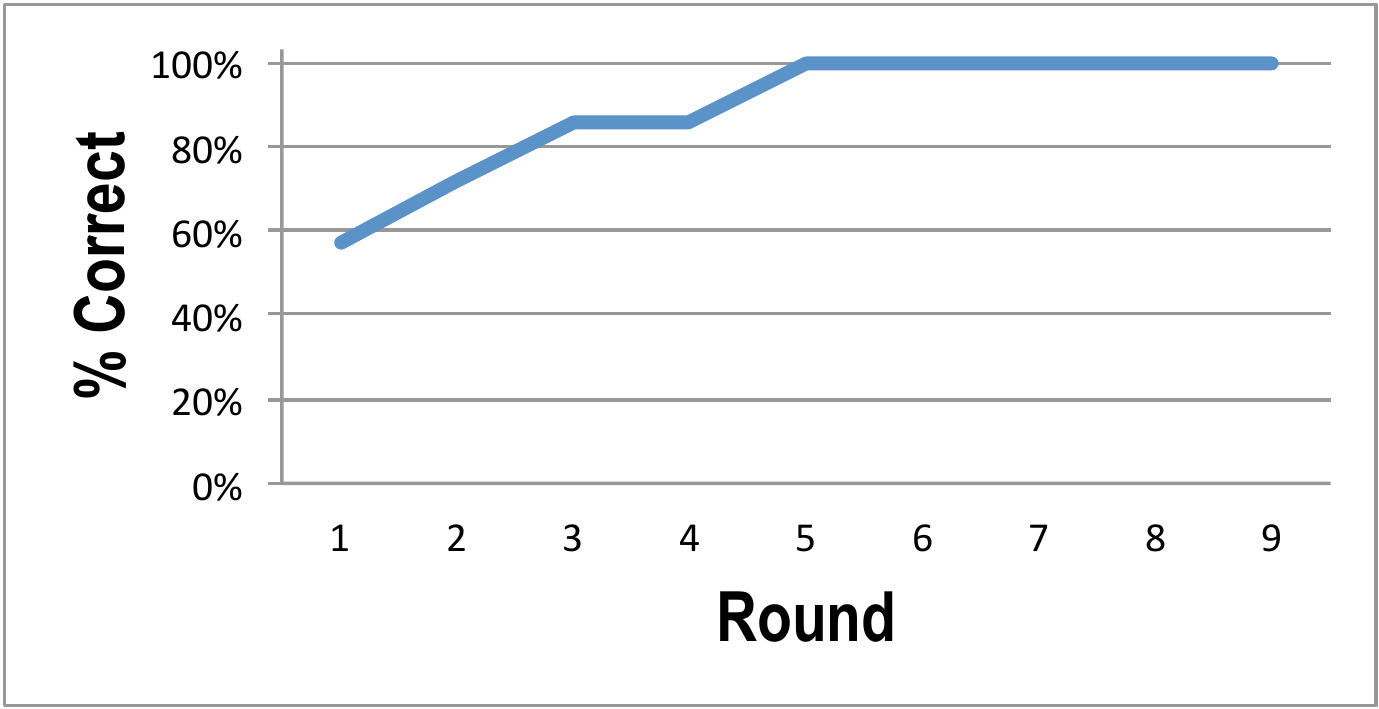}
\vspace{-1pc}
\caption{Average accuracy of the crowd decision in pressing the button to move to
		  the next round in our game in the 10-second trials.}
\label{fig:jsButtonsCorrect}
\end{figure}

\begin{figure}
\centering
\includegraphics[width=20.1pc]{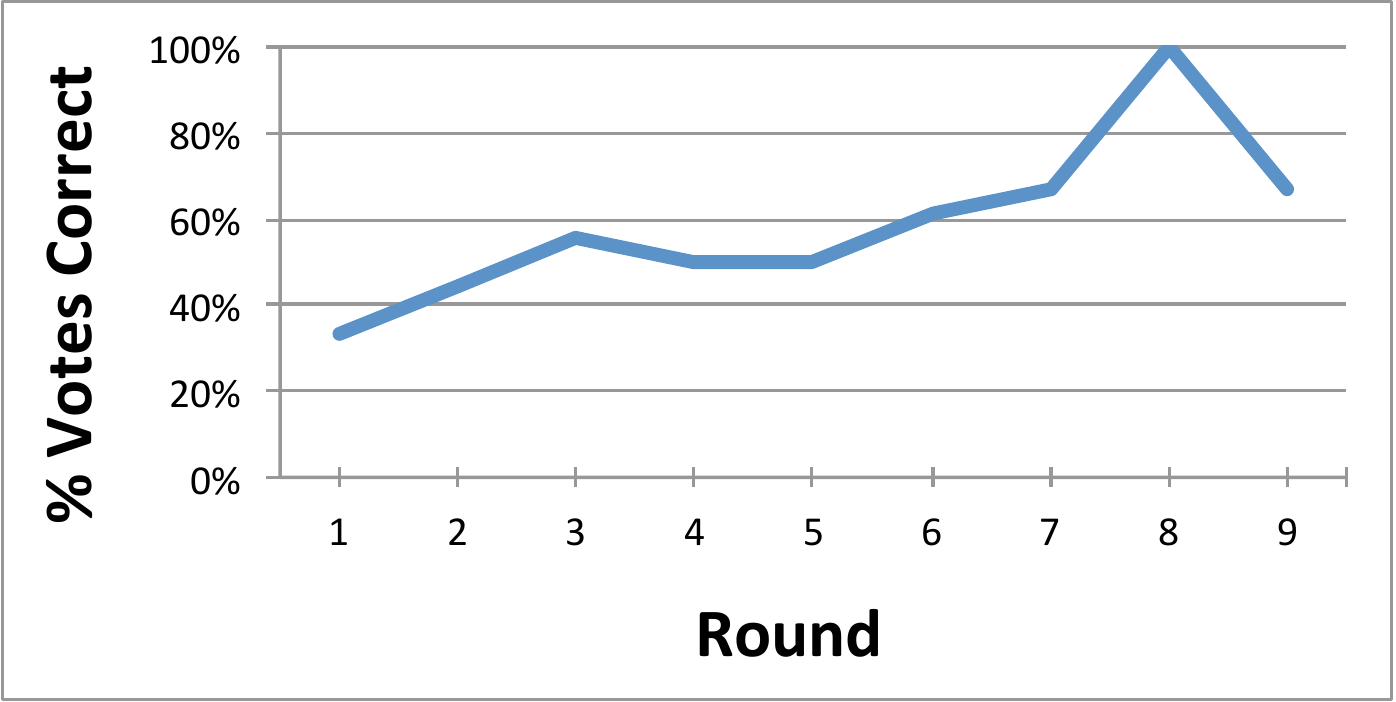}
\vspace{-1pc}
\caption{Average percentage of voters pressing the correct button
		  in the 30-second trials. This shows that workers learned
		  the correct button to press by observing the crowd's actions
		  over time.}
\label{fig:jsButtonsAgree}
\end{figure}

\begin{figure*}
\centering
\includegraphics[width=40.3pc]{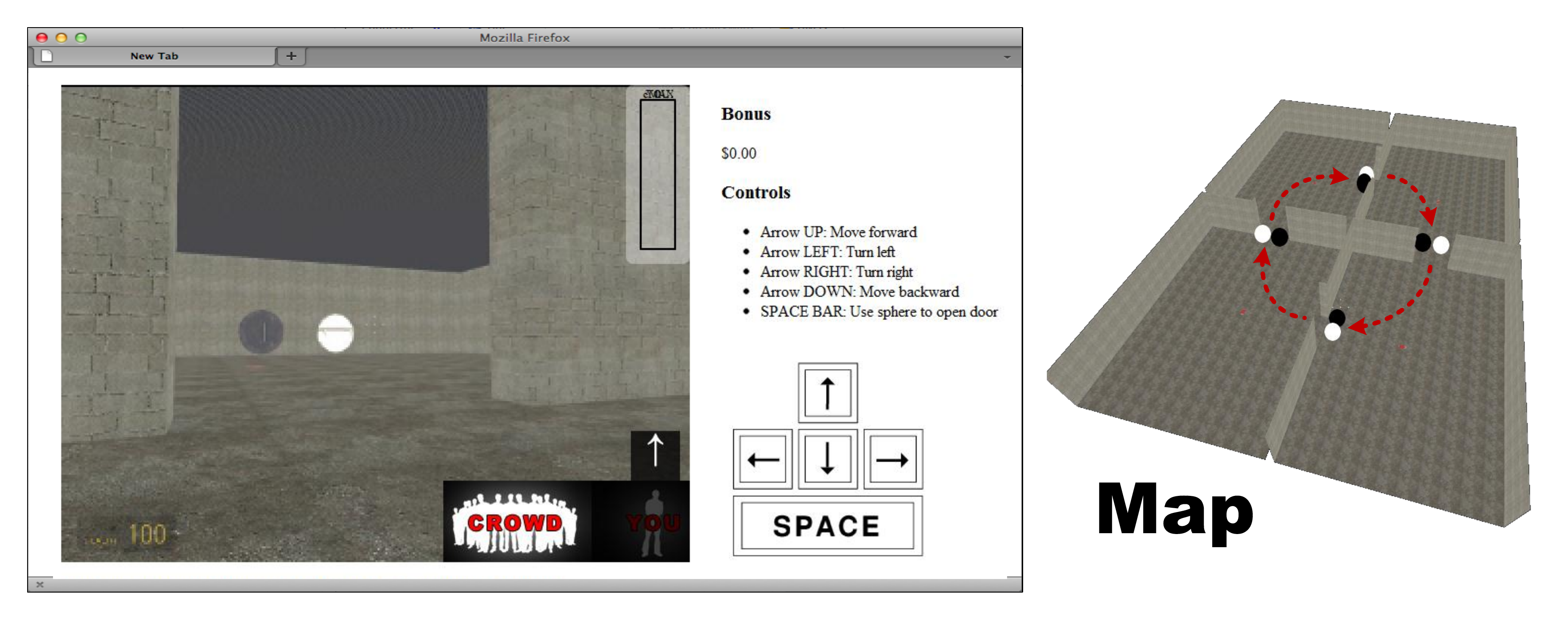}
\vspace{-1pc}
\caption{The worker interface for the continuous control task (left) and the
		  Half-Life 2 maze map that was used (right).}
 \label{fig:game-map}
\end{figure*}

\subsection{Legion}
Legion is a system which enables the crowd to collectively control a
single-user interface in real-time \cite{legion}. Individual workers
contribute input to a task as if they are independent of each other,
but the result of the group input is a single control stream that is
forwarded to the end-user's interface. This is done by recruiting a
set of workers to control the task, and presenting each worker with
an interface similar to a remote-desktop\cite{VNC}. Workers see a video
stream of the interface being controlled and their input is forwarded
to an \emph{input mediator} that combines the collective input into
a single control stream. One of the primary benefits of Legion is
its ability to let the crowd control an existing single-user interface, thus
eliminating the need for tasks to be broken down into pieces and
customized specifically for completion by the crowd.

While removing the need to explicitly divide tasks into
smaller pieces saves requesters effort, it also means
that many tasks will have an unknown length and
complexity a priori, making it difficult to properly
motivate workers.
HiveMind \cite{hivemind} is a crowdsourcing model that
uses game theoretic mechanisms to motivate workers to complete
tasks as quickly as they can accurately. The model uses a voting
system that encourages workers to aggregate their responses
via agreement with previously proposed answers. HiveMind can
successfully motivates workers to complete work and segment
tasks in continuous real-time tasks by using a reputation mechanism
to encourage workers to answer honestly, and by learning
which workers can be used as consistent representative leaders
for Legion.

\subsubsection{A single crowd-agent}
The crowd-agent model introduced by Legion aims to use the
collective not as a group of workers of varying reliability, but
as a single dependable agent. This approach
enables new styles of interaction with the crowd and can exhibit attributes
beyond that of a traditional single worker. For example, the crowd
agent can continue to work on a task indefinitely by recruiting new
workers, whereas a single individual would eventually have to rest.
The crowd can also complete tasks in parallel, even if the overall output
is the same as one worker. This model closely parallels that seen
in large organizations such as corporations, where the entity itself
acts as an individual, but can produce and process in parallel.

\subsubsection{Memory}
While this agent model of the crowd is promising,
it is important that the resulting crowd agent
retains all (or most) of the properties of a single worker.
One key property is memory. When a single user is
performing a task, they naturally have a history of
previous actions and states stored in memory that they use to inform future decisions.
However, the crowd agent is composed
of a number of workers, and has continuous turnover of these members --
meaning that eventually the agent will continue to work past the point
where any of the original members remain.
In this case, we investigate whether or not the crowd is
able to retain information as a collective, even if the workers
who directly observed an event are not available. We propose that
due to the crowd's ability to react to both the environment and each other in
real-time, the type of organizational learning seen in existing social groups
will be possible within the crowd. This style of learning can also be applied
to systems that share state information with subsequent workers via
communication channels such as leaving comments about previous actions
or the voting feedback system used in HiveMind.

\begin{figure*}
\centering
\includegraphics[width=40.3pc]{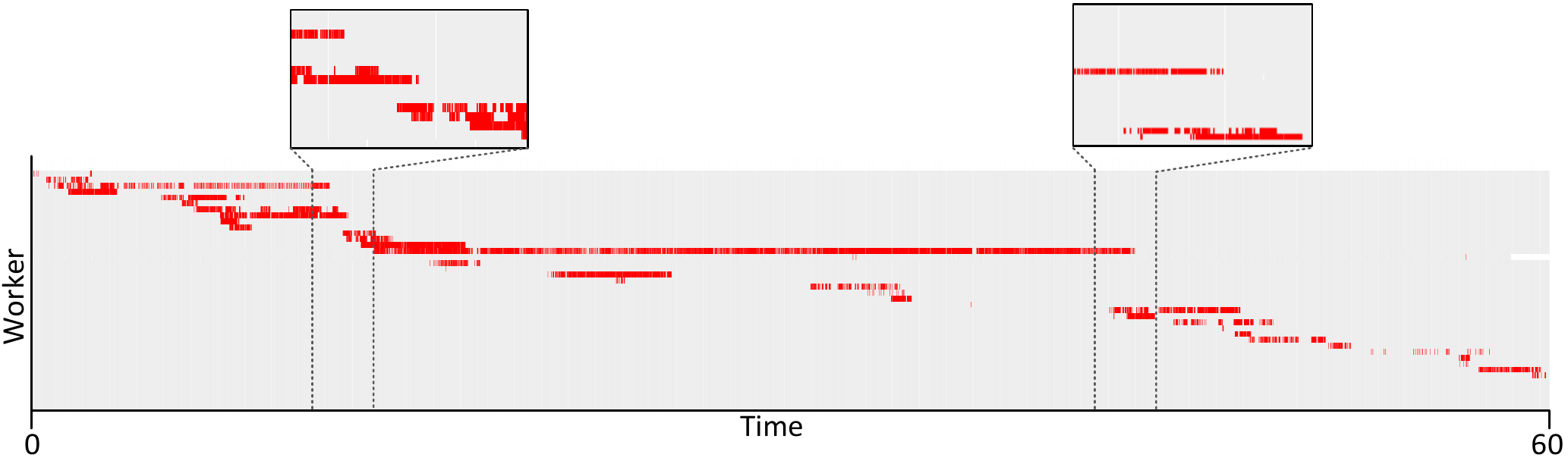}
\vspace{-1pc}
\caption{A visualization of the active workers throughout part of the task.
		  Each path is a worker, and the red markings indicate the worker
		  was active during that time interval.}
\label{fig:coverage}
\end{figure*}

\subsection{Experiments}
We ran two sets experiments to test the memory of continuous crowds. The
tests were setup as follows: An initial set of workers was directed
to interact with a specific object in the environment, then after a time
period the current set of workers was asked to repeat the last operation.
This allowed us to test the current crowd's ability to repeat the action even
with members who were not directly given the instructions.
Feedback was then presented to the crowd about the action taken. Workers
were told they got the answer correct regardless of their actual choice so
that workers wouldn't try to correct their answers during the next round, but
instead just continue to repeat the input.
We claim that the workers 
who begin the task after the directions were provided will be able to infer the
correct answer by viewing the collective actions of the rest of the crowd.


\subsubsection{Interface Element Selection}
First, we tested the ability of the crowd to remember simple selection data.
We provided workers with a Javascript game containing 10 numerical
keys to select from. The initial set of workers was told to keep pressing a
specific button, then we asked the additional workers who connected
after the beginning of the task to continue to re-enter the input that they
saw reflected by the interface for either 10 or 30 seconds before giving
any additional feedback to the crowd. Each test consisted
of ten demonstration points (rounds). The time delay was used to test how
the crowd's memory was effected by increase in time between demonstration
points (feedback), in the face of turn-over in the crowd.

Figures \ref{fig:jsButtonsCorrect} and \ref{fig:jsButtonsAgree} show the
outcomes of the 10 and 30 second button trials respectively.
Figure \ref{fig:jsButtonsCorrect} shows the 10 second trials promoted convergence to
a single final answer as the rounds progress. However, for the 30 second
trial this was not the case -- convergence to a single answer was much
quicker, generally happening after one or two rounds. Accordingly, there
was very little in terms of a long-term trend over the number of rounds.
Figure \ref{fig:jsButtonsAgree} shows that the crowd was still able to
demonstrate learning by consolidating their votes onto a single
answer over time. Thus, the crowd's agreement became stronger as
workers learned the correct button to press.

\subsubsection{The Continuous Worker}
Since work in cognitive science has shown that
the cognitive load of a task can impact learning \cite{sweller_cognitive_1988},
we are interested in showing crowd learning during a more realistic
task, not just in an isolated domain. We used a similar experimental setup,
but this time it was embedded in a navigation task. We used the video game
Half-Life 2 with a custom `maze' map as a test environment.
Workers were given two buttons to select from in a
room, both of which open the door to the subsequent room. The initial
set of workers were instructed to press a specific one at the beginning
of the task. Once the door is opened, workers must navigate their character
across the room to the next door. As shown in Figure \ref{fig:game-map},
these rooms form a cycle, so we could test for as many rounds as required.
We let the crowd control the navigation for 1 hour while the ability
of the crowd to retain knowledge of the task was recorded.

Our goal was to
demonstrate that the crowd-agent model was able to use organizational learning
to effectively act as a worker capable of performing a single task indefinitely. 
The crowd was able to complete this task, pressing the correct button (defined as the previous button)
over 95\% of the time, despite having several complete turnovers of workers.
Being able to use the crowd to control a continuous process for this length of
time shows the potential for much longer sessions. As long as crowds are made 
available to the system in sufficient numbers, this type of transfer learning can
continue indefinitely.


\subsection{Discussion}
Our experiments showed that Legion can support organizational learning
through implicit communication in the crowd. The results also revealed
distinct behaviors we did not expect for each of the trials.

The selection tests showed that in a very simple domain, workers
could learn a task. However, the tests also indicated a possible effect
from the amount of time between demonstration points on the
learning rate in terms of rounds. We believe this effect was due to
worker's attention being more focused when feedback is more scarce.
If this is supported by future tests, it may be possible to determine
tradeoffs between crowd size, task complexity, and attention.

The navigation test demonstrated that organizational learning within the
crowd-agent model works as expected. The crowd was able to correctly
select the button to press until near the very end of the task, at which
point, the crowd of unique workers we were recruiting became too sparse
and the number of connected workers hit zero temporarily, thus breaking
the chain of demonstrations and selecting the incorrect button.
However, using new tools for recruiting workers, and not
restricting workers to only those who  had not participated yet (which
was done to prevent workers with prior knowledge re-entering the test)
can help avoid such problems.

Figure \ref{fig:coverage} shows the activity of different crowd workers
connected to the task for a portion of the time. More workers with
experience connected to the task at once will make it easier to
influence crowd decisions and thus easier to demonstrate knowledge
to new workers. Here, we had a relatively small crowd, but one that
performed well on the task in part because new workers
seemed to be aware of their lack of domain knowledge, and held back
from participating until they had more confidence. In a few cases
this resulted in a single worker demonstrating the task to two or
more new workers. Encouraging this type of behavior may be required
to use smaller crowds reliably for tasks such as these.
HiveMind is able to motivate workers on continuous tasks, and encourage
them to abstain from answering in the case that they are unsure of the
correct answer, as is needed here. In future implementations,
we plan to incorporate such models to get better responses from the crowd. 

These results show that crowd agents can not only understand
natural language commands and remember past experiences like a single
person could, but can also work for an indefinite period of time, as
long as the crowd workers can continue to be recruited. Such agents
are also capable of outperforming the multiple (reliable) workers
recruited to complete a job sequentially since two workers would
not necessarily share a memory of the task without overlapping.
We paid approximately $\$3.60$/hour for work from
the crowd agent -- a price comparable to existing crowd systems.

\section{ANALYSIS}
 Crowd memory holds the promise of taking advantage of both enhanced skill sets of some workers, and pre-training workers to give them the general skills they will need to complete tasks quickly -- of critical importance when it comes to nearly real-time and real-time tasks.
 In the case of Legion for example, workers who do not understand how the interface works may provide little to no input if they enter in the middle of a task in completion. This hurts the task as a whole when only smaller crowds are available, as often workers would withhold input in order to learn or experiment with the controls, reducing the effective number of workers in the crowd.
 
 Several existing design considerations from general user interfaces become relevant to the design of crowdsourcing tasks in light of memory as a desired aspect of the crowd. We highlight a few of these below.
 
 \subsection{Learnability}
  Just as in any interface which we expect a user to learn to use effectively, we want the crowd to be able to learn our interfaces and tasks easily. One key aspect that must then be considered is \emph{learnability}, the difference here is that instead of a single user learning an interface or task, we have distributed the job to many workers in the crowd. There is a large literature in HCI describing considerations which should be made to improve learnability --- here we focus on a few of them in the context of distributed crowd learning:
  
  \subsubsection{Consistency} Consistency assists users in recalling what sets of knowledge to apply to a task. We may want whole interfaces to remain constant across tasks (such as in Legion), or we may only need specific functionality to mirror that which was learned in a previous portion of a task (such as a certain kind of text input box). Often visual consistency will be enough, but it is also important to make instructions and task descriptions similar so that workers with prior knowledge will be more inclined to accept a familiar task. It is possible that such considerations will improve workers' return rate because of their increased confidence in their ability to complete the task.
  
  \subsubsection{Explorable tasks} Making tasks explorable can enable much more complete knowledge of the domain by letting workers learn as much of the interface and task as they'd like. While this is not always ideal for a real task, since it may lead to situations where workers explore at the cost of quality input, it is vital to training tasks. In fact, one of the most beneficial uses of using crowd memory for training is being able to reduce the bad input (remove the noise) coming from good workers who spend some of their time exploring the interface. Current approaches rely on reducing the complexity of the interface to the point where most interaction options are clearly visible, in an attempt to minimize exploration (and its corresponding noise). This is particularly useful with crowds that are being compensated monetarily, and are effectively paid more to hurry through tasks, however, this limits the complexity and therefore functionality of the interfaces used in such tasks. If users are also paid to first learn the interface, they will be able to complete the actual task just as quickly, but more proficiently. Even with the ability to explore prior to a task, complexity must be managed carefully, as it still directly conflicts with its ability to be both learnable and memorable over time.
  
  \subsection{Motivation} 
  In order to spend the time learning a task in advance, we must motivate worker to explore the training task. This can be done using either the original payment scheme if the task is just a direct repeat of a real task (such as in the VizWiz case), or a modified payment scheme if the worker is expected to explore the interface in more depth --- though it may work for tasks which should be completed as quickly as possible, and have a known correct answer for the training task. Another option then is to pay workers to explore different aspects of the task by paying bonuses for using different functions.
  
  Since these training tasks can be designed to meet different specs than the actual task, mechanisms which may not work for a nearly real-time or real-time task could be made to work for training. For instance, making exploring the interface into a game may serve to better motivate workers to discover functionality, but may not naturally work to speed up the final task. This is alright as long as care is taken to ensure the key aspect of the interface that is learned is not the fact that additional time can be taken. The idea is to improve knowledge of the interface to later allow for more expert usage in higher importance domains.
  
 \subsection{Memorability}
 Despite workers' ability to remember certain aspects of a task, designing for memorability will help maximize this effect. Visual elements of a task can focus a worker's attention. Specifically designing for this will allow portions of an interface to be taught using training tasks that differ from the real task. In the case of a task which cannot be simulated or replayed (as is the case for Legion), this means that we can still take advantage of the benefits of pre-training, but in a piecewise fashion.


\section{FUTURE WORK}
  The use of crowd memory discussed in this paper provides the basis for research that changes some of the commonplace assumptions about the crowd and investigates taking advantage of the natural properties of human workers when it comes to learning. We can also leverage the fundamental differences between crowds and individuals to increase performance ever further.
 
 \subsection{Long Term Memory}
 While the ability to remember over the span of a day is an important initial step towards establishing the viability of crowd memory, the eventual goal is to enable workers to remember tasks across long periods of time such as days or weeks. This extended memory is more indicative of learning than memorization, and could lead to more in-depth training of the crowd over time.
 
 Preliminary tests on long term memory show that spans of several days are possible, but recall rates do fall over time as expected. To counteract this, we will focus on managing user assumptions and expectations. For example, users may expect an environment or configuration to change, while expecting the interface properties to remain the same. By making certain aspects of a task more salient, users can be subtly encouraged to memorize specific information for later use.
 
 \subsection{Explicit Communication}
 What we have presented in this paper focuses exclusively on passive and implicit communication, however, tasks can be designed to include explicit communication channels between workers in the several ways. We discuss a few such ways below:
 
 \ $\bullet$ \textbf{Instant messengers} can enable users to collaborate in a real-time task. As such, the organizational learning we explore in this paper could be used in cases where decision points do not exist or are not clearly observable. Workers are motivated to share their knowledge with others because it increases the likelihood of the task being completed correctly, and the group as a whole being rewarded. To account for the possibility of malicious users, different strategies can be used to moderate the conversation, such as restricting limiting the ability to add comments to special classes or workers or using rate limits to limit unproductive messages. History logs in which workers leave messages to both current and future workers can be used to extend this model and allow for long term communication. We expect this will take the form of a ranked forum, with more popular and currently relevant ideas bubbling to the top as workers promote or demote them.
 
 \ $\bullet$ \textbf{Automatic recording} of key events at certain inputs or at fixed intervals, such as screen shots or sped up video, could be used to quickly pass knowledge of what has been done in the task so far to new workers as they connect. The uses of communication amongst workers who are jointly completing a task clearly enables more than just knowledge sharing, and can also be used to collaborate by jointly deriving courses of action for accomplishing a task.
  
  \subsection{Complex Knowledge}
 We would also like to explore embedding more complex knowledge into the crowd and identify properties which are important when trying to prepare a crowd for a task in advance. Eventually this could lead to situations where the crowd can be trained in a specific area of knowledge prior to a task occurring. They will then be able to respond quicker and more accurately without needing an explicit certificate process as is often the case currently. The idea of crowd memory as a means of storage goes beyond just storing explicit knowledge or prior experience. It can include more deep information such as understanding of a domain, idea, or concept. 
 This can be used to solve problems that rely not only the ability to solve problems, but also recall available resources.
 
 \subsection{Continuous Learning}
  We can then merge long term memory, between-task recall, the ability to recruit large crowds, to enable continuous learning over longer periods of time. For instance, if we can re-recruit several previous workers, then they can learn knowledge they missed from the current crowd, and add the knowledge they had of their previous session. By maintaining a set of overlapping workers across multiple days, it maybe possible to continually grow the knowledge base of the crowd agent, making it more effective as time goes on. Work on eliciting this behavior could ultimately enable behaviors such as collective memory \cite{collectivememory} to manifest in the crowd.
 
 \subsection{Enhanced Cognitive Load}
 We have shown that the crowd agent has properties a single worker does not -- here, the ability to work continuously for an indefinite period of time. Work span is not the only property that the crowd has that an individual does not (i.e. the ability to truly work in parallel). We are currently investigating the possibility that the crowd's maximum cognitive load when working as a single agent can exceed that of a single individual, and in fact grows as a function of the crowd's size. This can be viewed as using the crowd as a shared distributed memory system. Since cognitive load is related to the total working memory and processing capacity of the agent, the extended working memory and parallel solving ability of multiple workers may allow the total maximum cognitive load of the crowd agent to be greater than any individual worker.
 
 
 \subsection{Enhanced Memory}
 The memory of a crowd agent is likewise extensible based on the properties of the crowd. Since workers will not remember identical portions of segments of informations, we are investigating whether it's possible to use multiple workers to reconstruct a greater portion of information than a single worker could. For example, given a large set of numbers with crowd workers each told to remember as many as possible, we expect the collective to remember more of the set than any single member. We intend to test this using a test similar to the example given here.

\section{CONCLUSION}
This paper explores crowd memory and learning the context of human computation. Through several experiments on Amazon's Mechanical Turk service we have shown the the crowd can learn quickly, and retain knowledge after a task has ended. This promotes the idea of designing tasks which stray from the traditional assumption of an untrained worker.

 We have also showed that crowds are able to train new workers in continuous tasks through a process similar to organizational learning, helping to offset problems introduced by high turnover rates in such tasks. This could allow the crowd-agent paradigm introduced by Legion, in which multiple workers act as a single, more reliable worker, to be used where traditional workers cannot, such as perpetual tasks which require some memory of previous actions. We also plan to explore extensions to this concept that will show the crowd's ability to use collective memory to more effectively complete tasks as an agent that are difficult even for a single user.


\bibliography{totalrecall}

%

\end{document}